# HIPI Kullanarak Çok sayıda Raster Uydu Görüntüsünün Dağıtık Mimaride Vektörleştirilmesi

# Vectorization of Large Amounts of Raster Satellite Images in a Distributed Architecture Using HIPI

Süleyman Eken, Eray Aydın, Ahmet Sayar

Bilgisayar Mühendisliği, Kocaeli Üniversitesi, Umuttepe Kampüsü, İzmit, 41380, Türkiye
{suleyman.eken, ahmet.sayar}@kocaeli.edu.tr, eray.aydin@yandex.com.tr

*Özetçe*—Vektörleştirme, piksel formdaki raster görüntülerin çizgi segmentlerine, çoklu çizgi ve poligon formuna dönüştürmeye odaklanmaktadır. Kaynak ve performans problemleri dikkate alınarak çok sayıda raster görüntüyü vektörleştirmek için MapReduce tabanlı HIPI görüntü işleme dağıtık büyük veri arayüzü kullanılmıştır. Apache Hadoop bu çatının çekirdeğinde yer almaktadır. Böyle bir sistemi gerçeklemek için ilk olarak map fonksiyonu ile girdi ve çıktı formatları tanımlanmıştır. Bu çalışmada, map fonksiyonları raster görüntüyü vektöre çevirmektedir. Reduce fonksiyonlarına vektörizasyon için ihtiyaç duyulmamıştır. Bant genişliği probleminin negatif etkilerini azaltarak dağıtık hesaplamada daha iyi sonuç almak için raster görüntülerin vektör temsilleri oluşturulmuş ve yatay ölçeklenebilirlik analizleri yapılmıştır.

*Anahtar Kelimeler* — vektörleştirme, raster-vektör görüntü, MapReduce, HIPI, OpenCV, LandSat 8

*Abstract*—Vectorization processes focus on grouping pixels of a raster image into raw line segments, and forming lines, polylines or polygons. To vectorize massive raster images regarding resource and performance problems, we use a distributed HIPI image processing interface based on MapReduce approach. Apache Hadoop is placed at the core of the framework. To realize such a system, we first define mapper function, and then its input and output formats. In this paper, mappers convert raster mosaics into vector counterparts. Reducer functions are not needed for vectorization. Vector representations of the raster images is expected to give better performance in distributed computations by reducing the negative effects of bandwidth problem and horizontal scalability analysis is done.

*Keywords* — vectorization, raster-vector image, MapReduce, HIPI, OpenCV, LandSat 8

## I. GİRİŞ

Uzaktan algılama; geomatik, enformatik, geofizik, harita mühendisliği, cartografya, vb. bilim alanlarında çeşitli çalışmalara (nesne çıkarımı/takibi/tanıma, uzamsal analizler, topolojik analizler, askeri ve sivil simulasyon uygulamaları, artırılmış gerçeklik uygulamaları vb.) yardımcı olmaktadır [1].

Uydu görüntülerinden nesne çıkarımı (yol, köprü, bina vb.) da uzaktan algılama görüntü işleme araştırma alanlarından biridir [2][3]. Literatürde sulu alanlar içinde kalan bölgelerden kara parçalarının çıkarılması (ada gibi) [4][5] veya kara parçaları üzerinde kalan sulu kısımların çıkarılması (göl, akarsu vs. gibi) nesne çıkarımı alt alanları içerisinde incelenmektedir [6][8]. Bu çalışmada ada gibi nesnelerin kenar bulma tabanlı olarak çıkarılması ve vektörizasyonu incelenmektedir. Vektör hale getirilen görüntüler üzerinde uzamsal analizler yapmak ve bu görüntüleri makineler arası efektif bir şekilde transfer etmek mümkündür [9]. Uydu sensörlerinin uzamsal, spektral ve zamansal çözünürlüklerinin hızlı bir şekilde artması ile birlikte uydu görüntülerinin büyüklüğü ve karmaşıklığı da üstel olarak artmıştır. Görüntülerin boyutunun çok büyük olmasından dolayı onları işlemek için ihtiyaç duyulan işlem-gücü ve bellek büyüklüğü gereksinimi de artmaktadır. Bu da, işlemlerin tek makinada (merkezi) gerçeklenebilmesini imkansız kılmaktadır. Son zamanlarda bu tür büyük verilerin dağıtık ve paralel olarak işlenmesine olanak tanıyan Hadoop [10], Spark [11], H2O [12] gibi açık kaynak yazılımlar ve altyapılar araştırmacılar tarafından sunulmuştur. Bu çalışmadaki amacımız çok sayıda uydu görüntüsünün dağıtık olarak işlenebilmesini sağlayacak bir hesaplama mimarisinin gerçeklenmesidir. Mimarinin temeli Eşle/indirge (Map/Reduce) paradigması dayanmaktadır. Eşle/indirge yönteminin yaygın kullanımına bakıldığında genelde web tabanlı verilerin dağıtık olarak kaydedilmesi ve kaydedilen veriler üzerinde aramalar yapılabilmesini sağlamak amaçlıdır. Bundan dolayı dosya giriş-çıkış formatlarının uydu görüntülerini işleyecek şekilde değiştirilmesi gerekmektedir. Uydu görüntülerini paralel olarak işleyebilmek için HIPI'ye (Hadoop Görüntü İşleme Arayüzü-Hadoop Image Processing Interface) fonksiyonaliteler eklenmiştir.

Çalışmanın geri kalanı şu şekilde organize edilmiştir. 2 bölümde, MapReduce programlama paradigması, HIPI arayüzü gibi temel anlatımlar sunulmuştur. 3. bölümde, uydu görüntülerinin dağıtık olarak işlenmesi ile ilgili literatürdeki çalışmalar sunulmuştur. 4. bölümde, uydu görüntülerinin vektörleştirilmesi metodolojisi verilmiştir. 5. bölümde HIPI kullanılarak gerçekleştirilen birtakım testler ve analizler verilip sonuçlar değerlendirilmiştir. Son kısımda ise sonuçlar sunulup gelecek çalışmalardan bahsedilmiştir.

## II. TEMEL ANLATIMLAR

### A. MapReduce Programlama Modeli

MapReduce; dağıtık mimari üzerinde çok büyük verilerin kolay bir şekilde analiz edilebilmesini sağlayan bir yapıdır. 2004 yılında Google tarafından duyurulan bu sistem, daha eski yıllarda geliştirilen map ve reduce fonksiyonlarından esinlenilmiştir. Veriler işlenirken bu iki fonksiyon kullanılır. Map aşamasında ana düğüm verileri alıp daha ufak parçalara ayırarak işçi düğümlere dağıtır. İşçi düğümler işi tamamladıkça sonucu ana düğüme gönderir. Reduce aşamasında ise tamamlanan işler, işin mantığına göre birleştirilerek sonuç elde edilir. Kısacası; map aşamasında analiz edilen veri içerisinden almak istediğimiz veriler çekilir. Reduce aşamasında ise çekilen bu veri üzerinde istediğimiz analiz gerçekleşir. Apache Hadoop, datığık olarak veri işleme ve analiz imkanı sunan açık kaynaklı yazılımdır *[13]*. 2011 yılında yayınlanan Apache Hadoop, Java programlama dili ile yazılmıştır. Hadoop dört ana modülden oluşur. Ancak bu modüllerin dışında da datığık sistemler üzerinde çalışan birçok yazılımıda destekler. Hadoop'un ana modüllerinden biri olan MapReduce ile veri işleme ve analizi gerçekleştirilmektedir. MapReduce içerik olarak iki tip özel metodu barındırır. Bu metodlar map ve reduce metodlarıdır. Map metodu filtreleme ve sıralama, reduce metodu ise işlemsel toplamayı sağlamaktadır. MapReduce uygulamaları yapısal olarak metin tabanlı veri işleme ve analizine uygundur; ancak diğer tip verileri işlemek için map ve reduce metotlarında bir takım giriş ve çıkış formatlarının ayarlanması gereksinimi vardır. HIPI görüntü işleme kütüphanesi MapReduce paradigmasında imge dosyalarını işlemek için OpenCV kütüphanesiyle birlikte çalışan bir mimariyi sunmaktadır.

### B. HIPI Görüntü İşleme Arayüzü

HIPI, Virginia Üniversitesinin Hadoop mimarisini temel alarak büyük boyuttaki görüntü verilerininin dağıtık olarak işlenmesini sağlayan bir altyapıdır [14]. Normal platformlarda büyük ölçekli görüntülerin işlenmesi oldukça zordur; ancak HIPI ile yapılan uygulamalar kullanıcıların büyük ölçekli görüntüleri kolayca işlemelerini sağlamaktadır. Şekil 1'de HIPI mimarisi gösterilmektedir.

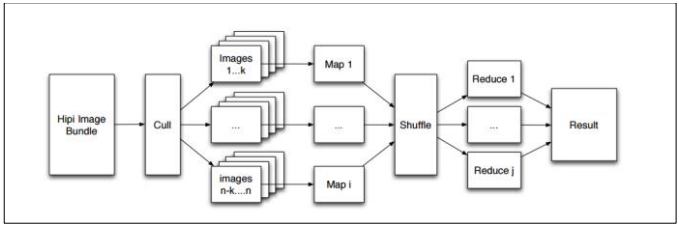

Şekil. 1. HIPI mimarisi [15]

HIPI mimarisinde bulunan modüller: (i) HIPI Image Bundle (HIB): HIPI mimarinin temel noktası bu kısımdır. Bu kısımda giriş olarak verilen birçok görüntü dağıtık dosya sisteminde tek bir HIB dosyası olarak tutulmaktadır. hibImport aracılığı ile birçok görüntüden HIB dosyası elde edilebilir. (ii) Culling Adımı: Hipi Image Bundle'daki görüntülerin birtakım görüntü özelliklerine göre filtrelenmesini sağlamaktadır. Bu filtre kullanıcının/programcının kararına bırakılmakla beraber piksel sayısı, dosya çeşidi, görüntü büyüklüğü vb. gibi filtreler olabilmektedir. (iii) Map ve Reduce Adımları: Filtrelerden geçirilen input dosyaları küçük image dosyalarına ayrıldıktan sonra kullanıcının kurduğu algoritmalara göre Map ve Reduce metodlarına gönderilerek işlenmeyi sağlamaktadır.

## III. İLGİLİ ÇALIŞMALAR

Bu bölümde ilk olarak Hadoop MapReduce kullanarak uydu görüntüleri üzerinde literatürde var olan çalışmalar özetlenecek daha sonra HIPI arayüzü ile yapılan işlere yer verilecektir. Winslett ve arkadaşları [16], eşle/indirge programlama modeli kullanarak mekânsal veri setlerinin paralel işlemesi için bir çalışma yapmışlardır. Çalışmada, eşle/indirge çatısının vektör ve raster veri gösterimlerinin paralel olarak işlenmesi için nasıl kullanılacağı açıklanmıştır. Golpayegani ve Halem [17], Lv ve arkadaşları [18], Hadoop'un eşle/indirge modelini kullanarak bazı uydu görüntü işleme algoritmalarını gerçeklediler; fakat Hadoop'ta görüntüleri ham olarak kullanmadan önce text formatına sonra da ikili formata çevirdiler. Görüntüleri ham olarak kullanmadıkları için bu ön işlem çok hesaplama zamanı almıştır. Ermias da [19] çalışmasında eşle/indirge temelli büyük boyutlu uydu görüntülerinin işlenmesini sunmuş; Sobel, Laplacian ve Canny gibi kenar bulma algoritmaları üzerinde durum çalışması yapmıştır. Junfeng ve arkadaşları [20] ise Hadoop'ta uzaktan algılama görüntülerinin nasıl yönetileceğinden bahsettikten sonra statik ve dinamik bir web harita servisinin nasıl tasarlanacağını anlatmışlardır. Xuhui ve arkadaşları [21] internet üzerinde GIS görüntülerini kaydeden ve kullanıcılara bu görüntüler üzerinde arama yapma imkânı veren Hadoop sunucularında küçük görüntü dosyalarının kaydından kaynaklı yavaşlığı gidermeye yönelik olarak dosyaların birleştirilerek HDFS'ye kaydını anlatan bir tekniği açıklamışlardır. Sarade ve arkadaşları [22] uydu görüntülerinden arazi örtüsü sınıflandırması için Hadoop hesaplama platformu önermişlerdir. Paylaşımlı bellekli (shared memory) sistemlerde birbirinin tamamıyla aynı olan çok sayıda işlemciden oluşur. Paylaşımlı bellekli paralel sistemler hem CPU üzerinde hem de GPU üzerinde gerçekleştirilebilmektedir. Ma ve arkadaşları [23] küme bilgisayarlardan (cluster/distributed memory) ziyade GPU programlama ile uydu görüntülerinin nasıl işlenebileceği hakkında bilgi vermişlerdir. Dong ve arkadaşları [24] büyük görüntü verilerini işlemek için hiyerarşik yapıda statitk ve dinamik görüntü bulutu işleme olmak üzere iki mekanizma içeren dağıtık bir alt yapı mimarisi önermişlerdir. Yaptığımız araştırmalara göre dağıtık uzaktan algılama görüntü işleme çalışmalarında düşük seviye görüntü işleme operasyonları (kenar bulma, gürültü azaltma-kaldırma), çok azında da orta seviye operasyonlar gerçekleştirilmiştir. Gerçekleştireceğimiz uzaktan algılama görüntülerinin birleştirilmesi ve nesne çıkarma çalışmasında yüksek seviyeli görüntü işleme operasyonlarının dağıtılması söz konusudur. Bu yönden de çalışma farklılık göstermektedir.

HIPI ile ilgili literatürde var olan çalışmalar sınırlıdır. Wilder ve arkadaşları [25] HIPI tüm görüntü formatlarını desteklemediğinden TIFF veya GeoTIFF formatındaki görüntüleri işleyebilmek için HIPI'ye eklenti geliştirmişlerdir.

LandSat 7 uydu görüntüleri üzerinde temel bileşen analizini test etmişlerdir. Basil ve arkadaşları [26] çok büyük çözünürlükteki cerrahi ameliyat videolarındaki kullanılan cerrahi aletleri tanımlamak için Hadoop'u kullanan bir cerrahi video analizi sistemi önermişlerdir. Video çerçeveleri HIB formatına çevrildikten sonra SIFT, SURF ve Haralick doku tanımlayıcıları vasıtasıyla cerrahi aletler tanımıştır. Akkoyunlu ve arkadaşları ise farklı iki veriseti üzerinden yüz bölgelerine ait biometrilerin saptaması ile ilgili performans incelemesi yapmışlardır [27].

## IV. UYDU GÖRÜNTÜLERİNİN VEKTÖRLEŞTİRİLMESİ METHODOLOJİSİ

Bu çalışmada Landsat-8 uydu görüntüleri; gri görüntüye dönüştürme, ikili görüntüye dönüştürme, morfolojik işlemler, ikili görüntüdeki boşluk-bölgelerin doldurulması, sınır piksellerinin takibine dayalı vektörleştirme işlemlerine tabi tutularak poligon-vektör modelleri oluşturulmuştur. Tüm işlemler Map metodunda gerçekleştirilmiştir. Ayrıca indirge (reduce) metodu kullanılmamıştır. Dağıtık vektörleştirme için HIPI arayüzü temel alınmıştır. HIPI arayüzünün FloatImage, HipiImageHeader ve ImageCodec sınıflarında değişiklikler yapılmıştır (Sayfa sayısının sınırlı olmasından kodlar eklenmemiştir).

Vektörleştirme için kullanılan yöntemlerin sözde kodu aşağıdaki gibidir :

Map metodu:

1. FloatImage'i opencv matrisine dönüştür (image to f)
2. matrisi grayscale 'e dönüştür (f to a)
3. a nın gray threshold 'unu hesapla. (thresh = graythresh(a))
4. a'yı binary 'e dönüştür. (a to bw)
5. bw'ye 300px threshold'u ile alan açma uygula
6. morfolojik 3x3 yapılandırma matrisi oluştur (se)
7. bw'ye se'yi kullanarak morfolojik görüntü kapama uygula
8. bw 'ye se'yi kullanarak morfolojik görüntü açma uygula
9. bw'deki bölge ve delikleri doldur (imfill)
10. bw'ye 10000px threshold 'u ile alan açma uygula
11. tüm dış noktalarını tespit et
12. tespit edilen noktaları vektöre ata (vector)
13. vector'u matrise yaz
14. matrisi FloatImage 'e dönüştür
15. FloatImage'i jpeg encoder ile hdfs'e kaydet

Görüntü işleme algoritmasının farklılığına göre indirge metodu veya zincirleme eşle-indirge yapıları da kullanılabilir. Takip eden bölümde vektörizasyonla ilgili performans testleri sunulmuştur.

## V. PERFORMANS DEĞERLENDİRMESİ

Çalışmamız için gerekli olan görüntüler LandSat uydularından elde edildi. LandSat programı National Aeronautics and Space Administration (NASA) ve U.S. Geological Survey (USGS) tarafından yönetilen bir uydu programıdır. Uyduların çekmiş olduğu görüntüler USGS'in çevrimiçi web sitesinden [28] ticari olmayan ve kar amacı gütmeyen araştırma faaliyetleri için ücretsiz sunulmaktadır. Site, kayıtlı kullanıcılarına farklı lokasyonlar ve farklı zaman aralıklarında uydu görüntülerini sunmaktadır.

Uydu görüntüleri görüntü parçalarından oluşur. Bu parçacıklara mozaik denir. Örnek bir mozaik görüntü Şekil 2'de sunulmuştur. Her bir mozaik, metadata denilen (veriyi tanımlayan veri) bir veri yapısıyla tanımlanır. Mozaiklerdeki görüntü piksel bilgileri ve görüntü işleme yöntemleri bu meta veri bilgileriyle birleştirilerek kullanılabilir bilgi (knowledge - wisdom) üretilebilmektedir. Şekil 3'te de bölüm IV'de verilen vektörleştirme işleminden geçmiş uydu görüntüsü sunulmuştur.

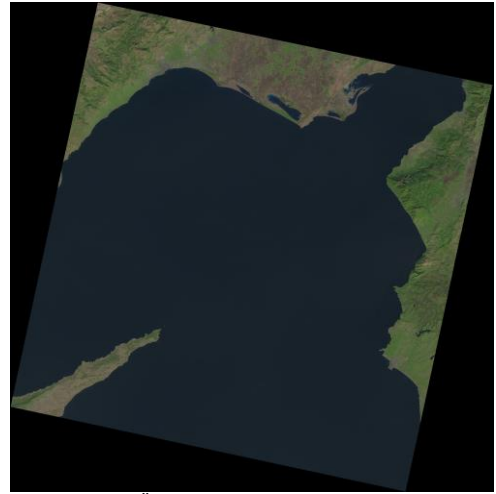

Şekil. 2. Örnek raster bir LandSat-8 mozaiği

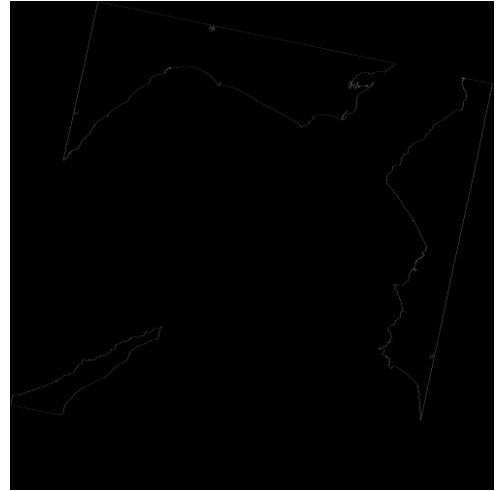

Şekil. 3. Vektörize edilmiş LandSat-8 mozaiği

Önerilen sistemi test etmek için 2 ve 4 bilgisayarlı bir Hewlett-Packard kümesi oluşturulmuştur. Her bir bilgisayar şu donanımsal özelliklere sahiptir: Intel(R) Core(TM) i7-3610QM CPU @ 2.30GHz, 8GB bellek, 160GB SATA disk. İşletim sistemi, Ubuntu olup kernel 3.13.0-37-generic kernel'ine sahiptir. Hadoop 2.6.0 ve Java 1.7.0 versiyonlarına sahiplerdir. Tablo I'de detaylı olarak vektörizasyon işleminin yatay ölçeklenebilirlik testleri gösterilmiştir. İki makineli bir Hadoop Cluster'ındaki koşum süresi ile dört makineli bir Hadoop Cluster'ındaki koşum süreleri belirtilmiştir. Ayrıca tabloda Matlab'ın aynı yöntemleri tek makinede ne kadar sürede çalıştırdığı da yer almaktadır. Tablodaki N değeri görüntü sayılarını göstermektedir. Her bir görüntü yaklaşık olarak 7000x7000 pikselden oluşmaktadır. Her üç senaryoda da görüntü sayısının artışıyla çalışma sürelerinin artışı gözlemlenmiştir. Dağıtık olmayan tek bir makina üzerindeki çalışma süreleri dağıtık mimarideki̇nden daha fazladır. Dağıtık mimaride paralel çalışan makina sayısının artışıyla çalışma sürelerinin düşüdü de gözlenmektedir.

TABLO I. VEKTÖRİZASYON ÖLÇEKLENEBİLİRLİK ANALİZİ

| Matlab için çalışma süresi (sn) | | İki makinada MapReduce için çalışma süresi (sn) | | Dört makinada MapReduce için çalışma süresi (sn) | |
|---|---|---|---|---|---|
| N=3 | N=20 | N=3 | N=20 | N=3 | N=20 |
| 67 | 582 | 56 | 315 | 54 | 261 |

## VI. SONUÇLAR VE GELECEK ÇALIŞMALAR

Görüntü işlemenin her alanda kullanımının yaygınlaşmasıyla çeşitli kaynaklardan üretilen çok büyük görüntü verilerinin depolanması, işlenmesi, indekslenmesi ve aranabilir hale getirilmesi gibi problemleri de beraberinde getirmiştir. Büyük görüntü verilerinin işlenmesindeki bu tür problemlerle başa çıkabilmek için gerek yazılım tabanlı gerekse donanım bazlı bir takım teknolojiler geliştirilmiştir. Biz de bu çalışmamızda büyük uydu görüntülerinin dağıtık bir mimaride HIPI arayüzü vasıtasıyla işlenebilmesi için yapılan geliştirilmelerden ve performans sonuçlarından bahsettik.

Vektör hale getirilen bu görüntülerden nesne çıkarmak amaçlı örme işleminin gerçekleştirilmesi [29] ve ayrıca bunlar üzerinde zaman-mekânsal analizler yapılması da gelecek çalışmalarımız içinde yer almaktadır.